\documentstyle[amssymb,epsf,cite,11pt]{article}

\def\baselinestretch{1.3}

\newcommand{\be}{\begin{equation}}
\newcommand{\ee}{\end{equation}}

\begin{document}

\title{\bf Multifractality and Laplace spectrum of horizontal visibility graphs constructed from fractional Brownian motions }

\author{ Zu-Guo Yu$^{1,2}$\thanks{
  Corresponding author, email: yuzg1970@yahoo.com}, Huan Zhang$^{1}$, Da-Wen Huang$^{1}$, Yong Lin$^{3}$ and Vo Anh$^2$\\
{\small$^1$Hunan Key Laboratory for Computation and Simulation in
Science and Engineering and }\\
{\small Key Laboratory of Intelligent Computing and Information
Processing of Ministry of Education,}\\
{\small Xiangtan University, Xiangtan,  Hunan 411105, China.}\\
{\small $^{2}$School of Mathematical Sciences, Queensland University of Technology,}\\
{\small GPO Box 2434, Brisbane, Q4001, Australia.}\\
{\small $^{3}$Department of Mathematics, School of Information, Remin University of China, Beijing 100872, China.} 
}
\date{}
\maketitle

\begin{abstract}
Many studies have shown that additional information can be gained on time series by investigating their associated complex networks. In this work, we investigate the multifractal property and Laplace spectrum of the horizontal visibility graphs (HVGs) constructed from fractional Brownian motions. We aim to identify via simulation and curve fitting the form of these properties in terms of the Hurst index $H$. First, we use the sandbox algorithm to study the multifractality of these HVGs. It is found that multifractality exists in these HVGs. We find that the average fractal dimension $\langle D(0)\rangle$ of HVGs approximately satisfies the prominent linear formula $\langle D(0)\rangle = 2 - H$; while the average information dimension $\langle D(1)\rangle$ and average correlation dimension $\langle D(2)\rangle$ are all approximately bi-linear functions of $H$ when $H\ge 0.15$. Then, we calculate the spectrum and energy for the general Laplacian operator and normalized Laplacian operator of these HVGs. We find that, for the general Laplacian operator, the average logarithm of second-smallest eigenvalue $\langle \ln (u_2) \rangle$, the average logarithm of third-smallest eigenvalue $\langle \ln (u_3) \rangle$, and the average logarithm of maximum eigenvalue $\langle \ln (u_n) \rangle$ of these HVGs are approximately linear functions of $H$; while the average Laplacian energy $\langle E_{nL} \rangle$ is approximately a quadratic polynomial function of $H$. For the normalized Laplacian operator, $\langle \ln (u_2) \rangle$ and $\langle \ln (u_3) \rangle$ of these HVGs approximately satisfy linear functions of $H$; while $\langle \ln (u_n) \rangle$ and $\langle E_{nL} \rangle$ are approximately a 4th and cubic polynomial function of $H$ respectively.
\end{abstract}



\section{Introduction}
\ \ \ \ Complex network theory has become one of the most important developments in
statistical physics \cite{Albert02}. Many studies have shown that complex networks play an
important role in characterizing complicated dynamic systems in
nature and society \cite{Song2005}. Studies have shown
that complex network theory may be an effective method to extract
the information embedded in time series \cite{Zhang06,Xu08,Lacasa08,Luque09,Donner10,Donner12}.
The advancement of network theory provides us with a new perspective
to perform time series analysis \cite{Donner10,Donner12}.
Especially we can further understand the structural features and
dynamics of complex systems by studying the basic
topological properties of their networks. Researchers have proposed some algorithms
to construct different complex networks from time series \cite{Small09}, such as complex networks from
pseudoperiodic time series \cite{Zhang06}, visibility graphs (VG) \cite{Lacasa08} and horizontal visibility graphs (HVG) \cite{Luque09},
state space networks \cite{Li08}, recurrence networks
\cite{Donner10,Donner12,Marwan09}, nearest-neighbor networks
\cite{Xu08,Liu10} and complex networks based on phase space reconstruction \cite{Gao09} .

Among the aforementioned methods, the visibility algorithm proposed by Lacasa {\it et al.} \cite{Lacasa08} has attracted many applications
from diverse fields \cite{XieZhou11}, including stock market indices \cite{Ni2009,Qian2010}, human stride intervals \cite{Lacasa09},
occurrence of hurricanes in the United States \cite{Elsner2009}, foreign exchange rates \cite{Yang2009}, energy dissipation rates
in three-dimensional fully developed turbulence \cite{Liu2010}, human heartbeat dynamics \cite{Shao2010,Dong2010},
diagnostic EEG markers of Alzheimer's disease \cite{Ahmadlou2010}, and daily streamflow series \cite{Tang2010}.
A VG is obtained from the mapping of a time series into a network according to the visibility criterion \cite{Lacasa08,Lacasa09}:
Two arbitrary data points ($t_a$, $y_a$) and ($t_b$, $y_b$) in the time series have visibility, and consequently become two connected vertices (or nodes)
in the associated graph, if any other data point ($t_c$, $y_c$) such that $t_a < t_c < t_b$ fulfills
$$y_c <y_a +(y_b -y_a)\frac{t_c-t_a}{t_b-t_a}. $$

Time series is defined in the time domain and the discrete Fourier transform (DFT) is defined on the frequency domain, the VG is defined on the ``visibility domain''. The DFT decomposes a signal in a sum of vibration modes, the visibility algorithm decomposes a signal in a concatenation of graph¡¯s motifs, and the degree distribution simply makes a histogram of such ``geometric modes''. The visibility algorithm is a geometric (rather than integral) transform.

A preliminary analysis \cite{Lacasa08} has shown that the constructed VG inherits several
properties of the series in its structure. Thereby, periodic time
series convert into regular graphs, and random series into random
graphs. Moreover, fractal time series convert into scale-free
networks, enhancing the fact that a power-law degree distribution
of its graph is related to the fractality of the time series.
Then Luque \textit{et al.} \cite{Luque09} proposed the HVG which
is geometrically simpler and forms an analytically solvable version of VG.  The HVG has been used to study the daily solar X-ray brightness data \cite{YuAnh2012} and protein molecular dynamics \cite{ZhouLY2014} by our group.

Self-similar processes have been used to model fractal
phenomena in different fields, ranging from physics,
biology, economics to engineering \cite{Lacasa09}.
Fractional Brownian motion (fBm) is a stochastic processes defined by $dX_t= \mu(t,X_t)dt+ \sigma(t,X_t)dB_t^H$, where $\mu$ is the drift coefficient, $\sigma$ is the diffusion coefficient. $B_t^H$ is a Gaussian process, the index $H$ is called Hurst exponent with $0 < H < 1$ after the British hydrologist H. E. Hurst. Note that for $H = 1/2$ we get the standard Brownian motion (standard Wiener motion), which we shall further denote by $W_t$ \cite{Mandelbrot1983}. Variance of fBm is $\sigma^2t^{2H}$. Variance also corresponds to mean squared displacement \cite{Makse1996,Rybski2012}, $EX_t^2$. If $H = 1/2$, the diffusion process is called normal diffusion and the variance of fBm is $\sigma^2t$. This is the same as cases of Brownian motion and Brownian motion with constant drift, which means the mean squared displacement (MSD) of a particle is a linear function of time. If $1/2 < H < 1$, the diffusion process is called super-diffusion (Levy flight and geometric Brownian motion both belong to super-diffusion). If $0 < H < 1/2$, the diffusion process is called sub-diffusion. We can choose different diffusion process to model the data with various mean squared displacement \cite{Mandelbrot1983,Makse1996,Rybski2012}. Lacasa {\it et al.} \cite{Lacasa09} showed that
the VGs derived from generic fBm series are
scale-free, and proved that there exists a linear relation
between the Hurst exponent $H$ of the fBm and the exponent $\gamma$ of the power law degree distribution in the associated VG. The visibility algorithm thus provides
another method to compute the Hurst exponent
and characterize fBm. Xie and Zhou \cite{XieZhou11} studied the relationship
between the Hurst exponent of fBm and the
topological properties (clustering coefficient and fractal
dimension) of its converted HVG. Our group \cite{LiuYu14} studied the topological and fractal
properties of the recurrence networks constructed from fBms.

Based on the self-similarity of fractal geometry \cite{Mandelbrot1983,Feder1988,Falconer1997},
Song {\it et al.} \cite{Song2005} generalized the box-counting method and used it in the field of complex networks.
As a generalization of fractal analysis, the tool of multifractal analysis (MFA) has a better performance on characterizing the
complexity of complex networks in real applications. MFA has been widely applied in a variety of fields such as
financial modeling \cite{Canessa00,Anh00}, biological systems \cite{Yu01,Yu03,Yu04,Yu06}, and geophysical data
analysis \cite{Yu07,Yu09,Yu10,Yu14}. In recent years, MFA also has been successfully used in complex networks
and seems more powerful than fractal analysis. As a consequence of this trend,
some algorithms have been proposed to calculate the mass exponent $\tau(q)$ and then study the multifractal properties
of complex networks. Furuya and Yakubo \cite{Furuya11} proposed an improved compact-box-burning algorithm for MFA of complex networks based on the algorithm introduced by Song {\it et al.} \cite{Song07}, and applied it to show that some networks have a multifractal structure.
Almost at the same time, Wang {\it et al.} \cite{Wang12} proposed a modified fixed-size box-counting method to detect the multifractal behavior of some theoretical and real networks, including scale-free networks, small-world networks, random networks, and protein-protein interaction networks.
Li {\it et al.} \cite{LiYu14} improved the algorithm of Ref. \cite{Wang12} further and used it to investigate the multifractal properties of a family of fractal networks introduced by Gallos {\it et al.} \cite{Gallos07}.
Then Liu {\it et al.} \cite{LiuYu14} studied the fractal and multifractal properties of the recurrence networks constructed from fBms. Recently, Liu {\it et al.} \cite{LiuYu15} employed the sandbox (SB)
algorithm which was proposed by T\'{e}l et al. \cite{Tel1989} for MFA of complex networks. By comparing the numerical
results and the theoretical ones of some networks, it was shown
that the SB algorithm is the most effective, feasible and accurate
algorithm to study the multifractal behavior and calculate the
mass exponent of complex networks.

In another direction, spectral graph theory has a long history. One of the main goals in graph theory is to deduce the principal properties
and structure of a graph from its graph spectrum. The eigenvalues of Laplacian operator are closely related to almost all major
invariants of a graph, linking one extremal property to another \cite{Chung1997}. There is no question
that eigenvalues play a central role in our fundamental understanding of graphs \cite{Chung1997}. The study of graph eigenvalues
realizes increasingly rich connections with many other areas of mathematics. A particularly important development is the interaction
between spectral graph theory and differential geometry \cite{Chung1997,LinYau10}.

In this work, we investigate the multifractal property and Laplace spectrum of the HVGs
constructed from fBms. First, we use the SB algorithm employed by Liu {\it et al.} \cite{LiuYu15}
to study the multifractality of these HVGs. We
then calculate the spectrum \cite{Chung1997,LinYau10} and
energy \cite{Gutman06} for the general Laplacian operator and normalized Laplacian operator of these HVGs.
We aim to identify the functional forms of possible relationships between the Hurst index of the fBm and the
multifractal indices, Laplacian spectrum and energy of the associated HVG.

\section{Horizontal visibility graph of time series}

\ \ \ \ A graph (or network) is a collection of vertices or nodes, which denote the
elements of a system, and links or edges, which identify the relations or
interactions among these elements. A large number of real networks are
referred to as \emph{scale-free} because the probability distribution $P(k)$
of the number of links per node (also known as the degree distribution)
satisfies a power law $P(k)\sim k^{-\gamma }$ with the degree exponent $%
\gamma $ varying in the range $2<\gamma <3$ \cite{Albert99}.

Luque \textit{et al.} \cite{Luque09} proposed the horizontal visibility
graph (HVG) which are geometrically simpler and analytically
a solvable version of VG \cite{Lacasa08}. Given a time series $\{x_1, x_2, ... ,x_n\}$,
two arbitrary data points
$x_{i}$ and $x_{j}$ in the time series have horizontal visibility,
and consequently become two connected vertices (or nodes) in the associated
graph, if any other data point $x_{k}$ such that $i<k<j$ fulfils
$$
x_{i}, x_{j} > x_{k}.
$$
Thus a connected, unweighted network could be constructed based on
a time series and is called its horizontal visibility graph (HVG).
Two nodes $i$ and $j$ in the HVG are connected if one can draw a horizontal line in the time series joining $x_i$ and $x_j$ that does not intersect any intermediate data height. Given a time series, its HVG is always a subgraph of its associated VG.
Luque {\it et al.} \cite{Luque09} showed that the degree distribution of an HVG constructed from
any random series has an exponential form $P(k) =
(3/4)exp(-k\ln(3/2))$. Then Lacasa {\it et al.} \cite{Lacasa2010}
used the horizontal visibility algorithm to characterize and
distinguish between correlated, uncorrelated and chaotic
processes. They showed that horizontal visibility algorithm is able to distinguish chaotic series from independent and identically distributed (i.i.d.) theory without needs for additional techniques such as surrogate data or noise reduction methods \cite{Lacasa2010}.
 Xie and Zhou \cite{XieZhou11} studied the
relationship between the Hurst index of fBm and the topological properties (clustering coefficient and
fractal dimension) of its converted HVG. In this work, we investigate the multifractal property,
Laplace spectrum and energy of HVGs constructed from fBms.

\section{Sandbox algorithm for multifractal analysis of complex networks}

\ \ \ \ The fixed-size box-covering algorithm \cite{Halsey86} is well known as one of the most common and
important algorithms for MFA. For a given
measures $\mu$ with support set $E_0$ in a metric space, we consider
the following partition sum
\begin{equation}
Z_{\epsilon}(q) = \sum_{\mu(B)\neq 0}[\mu(B)]^{q},
\end{equation}
$q \in R$, where the sum runs over all different nonempty boxes
$B$ of a given size $\epsilon$ in a box covering of the support
set $E_0$. The mass exponents $\tau(q)$ of the measure $\mu$ can be defined as
\begin{equation}
\tau(q) = \lim_{\epsilon\rightarrow 0}\frac{\ln Z_{\epsilon}(q)}{\ln \epsilon}.
\end{equation}
Then the generalized fractal dimensions $D(q)$ of the measure $\mu$ are defined as
\begin{equation}
D(q) = \frac{\tau(q)}{q-1}, ~~~~~~\textrm{for}~~ q \neq 1,
\end{equation}
and
\begin{equation}
D(q) = \lim_{\epsilon\rightarrow0}\frac{Z_{1,\epsilon}}{\ln \epsilon}, ~~~\textrm{for}~~ q = 1,
\end{equation}
where $Z_{1,\epsilon} = \sum_{\mu(B)\neq0} \mu(B)\ln\mu(B)$. Linear regression of $[\ln Z_{\epsilon}(q)]/(q-1)$ against $\ln
\epsilon$ for $q \neq 1$ gives estimates of the
generalized fractal dimensions $D(q)$, and similarly a linear
regression of $Z_{1,\epsilon}$ against $\ln \epsilon$ for $q = 1$.
In particular, $D(0)$ is the box-counting dimension (or fractal
dimension), $D(1)$ is the information dimension, and $D(2)$ is
the correlation dimension. Usually the strength of the multifractality
can be measured by $\Delta D(q)=\max D(q)-\min D(q)$.

In a complex network, the measure $\mu$ of each box can be defined
as the ratio of the number of nodes covered by the box and the
total number of nodes in the entire network. In addition, we can
determine the multifractality of complex network by the shape of the
$\tau(q)$ or $D(q)$ curve. If $D(q)$ is a constant or $\tau(q)$ is
a straight line, the object is monofractal; on the other hand, if
$D(q)$ or $\tau(q)$ is convex, the object is multifractal.

The sandbox (SB) algorithm proposed by T\'{e}l {\it et al.}
\cite{Tel1989} is an extension of the box-counting algorithm
\cite{Halsey86}. The main idea of this sandbox algorithm is that
we can randomly select a point on the fractal object as the center
of a sandbox and then count the number of points in the sandbox.
The generalized fractal dimensions $D(q)$ are defined as
\begin{equation}
D_{q} = \lim_{r \rightarrow 0}\frac{\ln \langle[M(r)/M(0)]^{q-1}\rangle}{\ln(r/d)}\frac{1}{q-1},~~~~q \in R,
\end{equation}
where $M(r)$ is the number of points in a sandbox with a radius of $r$, $M(0)$ is the total number of points in the fractal
object. The brackets $\langle \cdot \rangle$ mean to take statistical average over randomly chosen centers of the sandboxes. In fact, the above equation can be rewritten as
\begin{equation}
\ln(\langle [M(r)]^{q-1} \rangle)\ \propto \
D(q)(q-1)\ln(r/d)+(q-1)\ln(M_{0}).
\end{equation}
So, in practice, we often estimate numerically the generalized
fractal dimensions $D(q)$ by performing a linear regression of
$\ln(\langle [M(r)]^{q-1} \rangle )$ against $(q-1)\ln(r/d)$; and
estimate numerically the mass exponents $\tau(q)$ by performing a
linear regression of $\ln(\langle [M(r)]^{q-1} \rangle )$ against
$\ln(r/d)$.

Recently, Liu {\it et al.} \cite{LiuYu15} proposed to employ the sandbox
algorithm for MFA of complex networks.
Before we use the following SB algorithm to perform MFA of a
network, we need to apply Floyd's algorithm \cite{Floyd62}
of Matlab-BGL toolbox \cite{Gleich} to calculate the shortest-path
distance matrix of this network according to its adjacency
matrix $A$. The SB algorithm for MFA of complex
networks \cite{LiuYu15} can be described as follows.

\begin{enumerate}

\item[(i)]  Initially, make sure all nodes in the entire network are not selected as a center of a sandbox.

\item[(ii)] Set the radius $r$ of the sandbox which will be used to cover the nodes in the range $r \in [1, d]$,
where $d$ is the diameter of the network.

\item[(iii)] Rearrange the nodes of the entire network into
random order. More specifically, in a random order, nodes which
will be selected as the center of a sandbox are randomly arrayed.

\item[(iv)] According to the size $n$ of networks, choose the
first 1000 nodes in a random order as the center of 1000
sandboxes, then search all the neighbor nodes by radius $r$ from
the center of each sandbox.

\item[(v)] Count the number of nodes in each sandbox of radius $r$, denote the number of nodes in each sandbox as $M(r)$.

\item[(vi)] Calculate the statistical average $\langle [M(r)]^{q-1} \rangle$ of $[M(r)]^{q-1}$ over all 1000 sandboxes of radius $r$.

\item[(vii)] For different values of $r$, repeat steps (ii) to (vi) to
calculate the statistical average $\langle [M(r)]^{q-1} \rangle$
and then use $\langle [M(r)]^{q-1} \rangle$ for linear regression.
\end{enumerate}

We need to choose an appropriate range of $r \in [r_{min},
r_{max}]$, then calculate the generalized fractal dimensions
$D(q)$ and the mass exponents $\tau(q)$ in this scaling range. In
our calculation, we perform a linear regression of $\ln(\langle
[M(r)]^{q-1} \rangle)$ against $\ln(r)$  and then choose the
slope as an approximation of the mass exponent $\tau(q)$ (the
process for estimating the generalized fractal dimensions $D(q)$ is
similar).

By comparing the numerical
results and the theoretical ones of some networks, Liu {\it et al.} \cite{LiuYu15} showed
that the SB algorithm is the most effective, feasible and accurate
algorithm to study the multifractal behavior and calculate the
mass exponents of complex networks. Hence we use the SB algorithm employed by Liu {\it et al.} \cite{LiuYu15}
to study the multifractality of the HVGs constructed from fBms in this work.

\section{Laplacian spectrum and energy of complex networks}

\ \ \ \ Suppose $G$ is a undirected graph with vertex set $V$ and
edge set $E$. The distance between two vertices is the minimum number of edges
to connect them; the diameter of $G$ is the maximum of all the distances of the graph \cite{LinYau10}.

Denote the adjacent matrix of the graph $G$ as $A=(a_{ij})_{n\times n}$, the degree of vertex $i$ as $d_i$. $T$ is diagonal
matrix of degrees, i.e.
\begin{equation}
T=\left(\begin{array}{cccc}
d_1    &0 &\cdots  & 0\\
0      &d_2     &\cdots & 0 \\
\vdots &\vdots  &\ddots & \vdots\\
0     & 0  &\cdots  & d_n
\end{array}\right)
\end{equation}
Define the operator $L=(b_{ij})_{n\times n}$, where
\begin{equation}
b_{ij}=
\left\{
\begin{array}{ccc}
~~~d_i,~~~if ~i= j\\
~~-1,~~~if~ a_{ij}=1\\
~0,~~~others
\end{array}\right.
\end{equation}
It is obvious that $L=T-A$. This operator $L$ is the general Laplace operator.
The normalized Laplace operator ${\cal L}$ is defined as ${\cal L}=(c_{ij})_{n\times n}$ \cite{Chung1997}, where
\begin{equation}
c_{ij}=
\left\{
\begin{array}{ccc}
~~~~~1,~~~if~ i= j\\
-\frac{1}{\sqrt{d_{i}d_j}},~~~if~ a_{ij}=1\\
~~~0,~~~~others
\end{array}\right.
\end{equation}
Denote
\begin{equation}
T^{-\frac{1}{2}}=\left(\begin{array}{cccc}
d_1^{-\frac{1}{2}}   & 0  &\cdots  &0\\
0      &d_2^{-\frac{1}{2}}     &\cdots & 0 \\
\vdots &\vdots  &\ddots & \vdots\\
0    & 0  &\cdots  & d_n^{-\frac{1}{2}}
\end{array}\right)
\end{equation}
where we set $d_i^{-1}=0$ when $d_i=0$.
It is seen that ${\cal L}=T^{-\frac{1}{2}}LT^{-\frac{1}{2}}$.
The spectrum consists of the eigenvalues of the general Laplacian operator $L$ and
normalized Laplacian operator ${\cal L}$ of the graph. It can be proved that the
smallest eigenvalue $u_1$ of the general Laplacian operator $L$ and
normalized Laplacian operator ${\cal L}$ of a connected graph is equal to 0 \cite{Chung1997,LinYau10}.
Usually, the second smallest eigenvalue $u_2$ and the maximum eigenvalue $u_n$ have particular meaning.
The second smallest eigenvalue $u_2=u_2-u_1$ (because $u_1=0$) is called the Laplacian spectral gap \cite{LinYau10}.
The second smallest eigenvalue $u_2$ and maximum eigenvalue $u_n$ are related to the synchronizability of complex
networks \cite{Atay2006}. Hence we pay more attention to the second-smallest eigenvalue $u_2$, the third-smallest
eigenvalue $u_3$, the average maximum eigenvalue $u_n$ of these two Laplacian operators of a graph in this work.

The Laplacian energy \cite{Gutman06} , $E_{nL}$, is defined as
\begin{equation}
E_{nL}=\Sigma_{i=1}^{n}|u_i-\frac{2m}{n}|.
\end{equation}
where $u_i$ is the $i$th eigenvalue of the general Laplacian operator $L$ (or
normalized Laplacian operator ${\cal L}$) of the graph, $n$ and $m$ are the numbers of vertices and edges
in the graph respectively.

\section{Results and discussion}

\ \ \ \ In this work, we use the Matlab command ``wfbm" to generate fBm time series of parameter $H$ ($0<H<1$) and
length $n$ following the wavelet-based algorithm proposed by Abry
and Sellan \cite{Abry1996}. We consider fBm time series with length $n = 10^{4}$ and
different Hurst indices $H$ ranging from 0.05 to 0.95 (the step
difference is 0.05). For each value of Hurst index $H$, we generate 100 fBm time series
with the same $H$, then we convert them into 100 HVGs.

   For each HVG, we calculate the $D(q)$ and $\tau(q)$ curves
using the SB algorithm. We calculate the $D(q)$ and $\tau(q)$ curves with $q$ ranging from -10 to 10
(the step difference is set to 1/3). After checking carefully many times with visual inspection, we find the best linear regression range of $r$ is $r\in(20, 72)$ in our setting.
Hence we set the range $r\in(20, 72)$ in our computations. We provide the linear regression to estimate $D(q)$ for a HVG converted from a fBm time series with Hurst index $H=0.4$ in Figure 1 as an example.

\begin{figure}[ptb]
\centerline{\epsfxsize=10cm \epsfbox{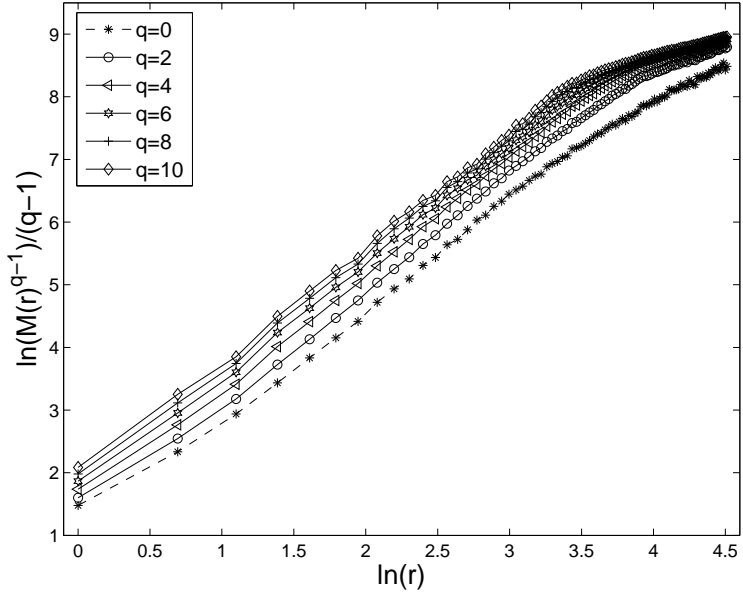}}
\caption{Linear regressions for calculating the generalized dimensions $D(q)$ of a HVG converted from a
fBm time series with Hurst index $H=0.4$.}
\end{figure}

In the following, the averages are taken over HVGs constructed from 100 time series of fBm with the same Hurst index $H$. We show the average $\langle \tau(q) \rangle$ curves and average $\langle D(q) \rangle$
curves in Figure 2. From Figure 2, we find that the $\langle \tau(q) \rangle$ and $\langle D(q) \rangle$ curves of HVGs are not straight lines, hence asserting that multifractality exists in these HVGs constructed from fBm series. We also find that the average multifractality of these HVGs becomes weaker, which is indicated by the value of $\langle \Delta
D(q)\rangle=\langle D(-10)-D(10)\rangle$, when the Hurst index of the given time
series increases, and the average multifractality is approximately a quadratic polynomial function of $H$ when $H\ge 0.1$ (as shown in the left panel of Figure 3).

\begin{figure}[ptb]
\centerline{\epsfxsize=6.5cm \epsfbox{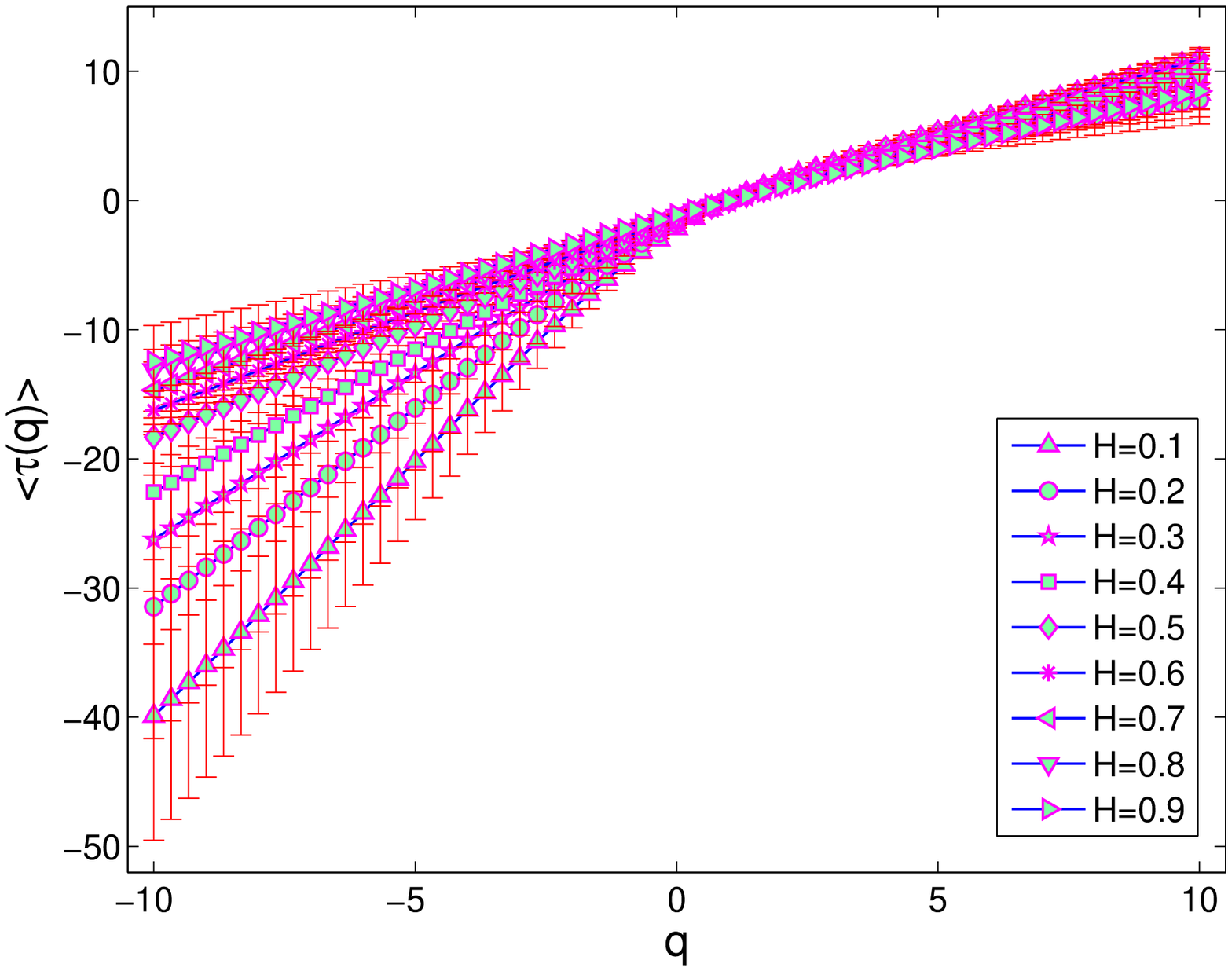}
\epsfxsize=6.5cm \epsfbox{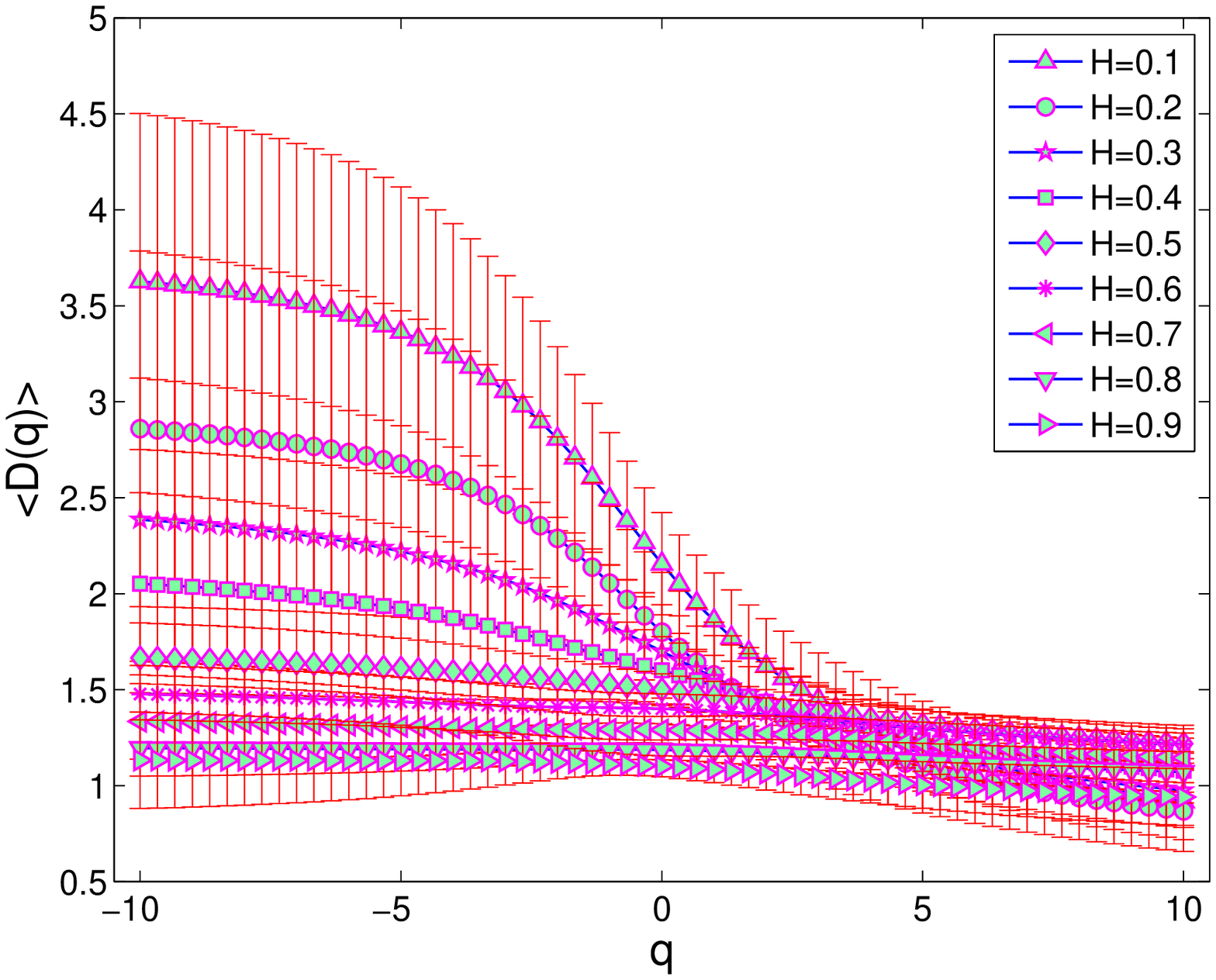}}
  \caption{The average $\langle \tau(q) \rangle$ {\bf (Left)} and $\langle D(q) \rangle$ {\bf (Right)} curves of the HVGs. Here the average
  is calculated from 100 realizations, and error bars are calculated by the standard errors.}
 \end{figure}

\begin{figure}[ptb]
\centerline{\epsfxsize=6.5cm \epsfbox{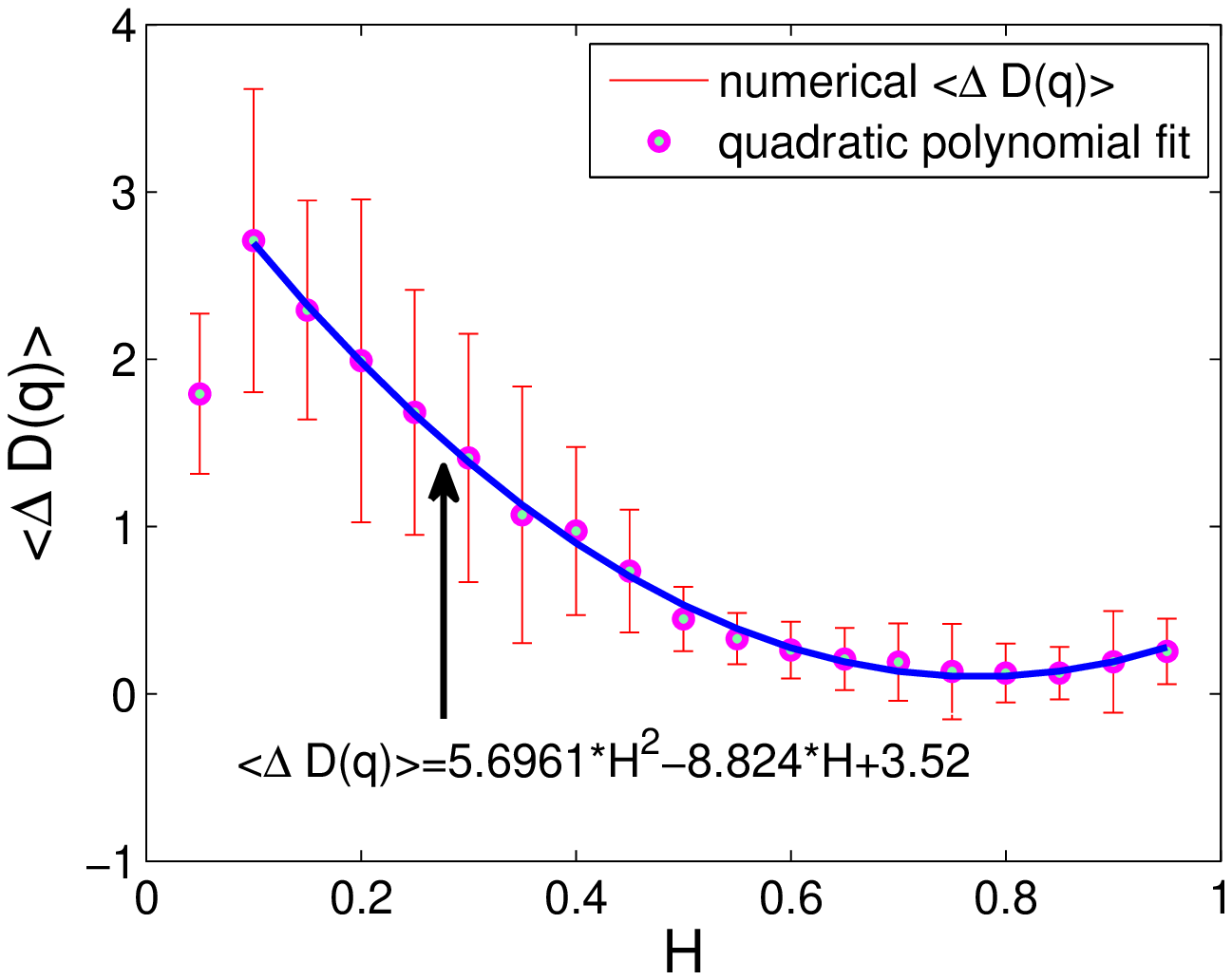}
\epsfxsize=6.5cm \epsfbox{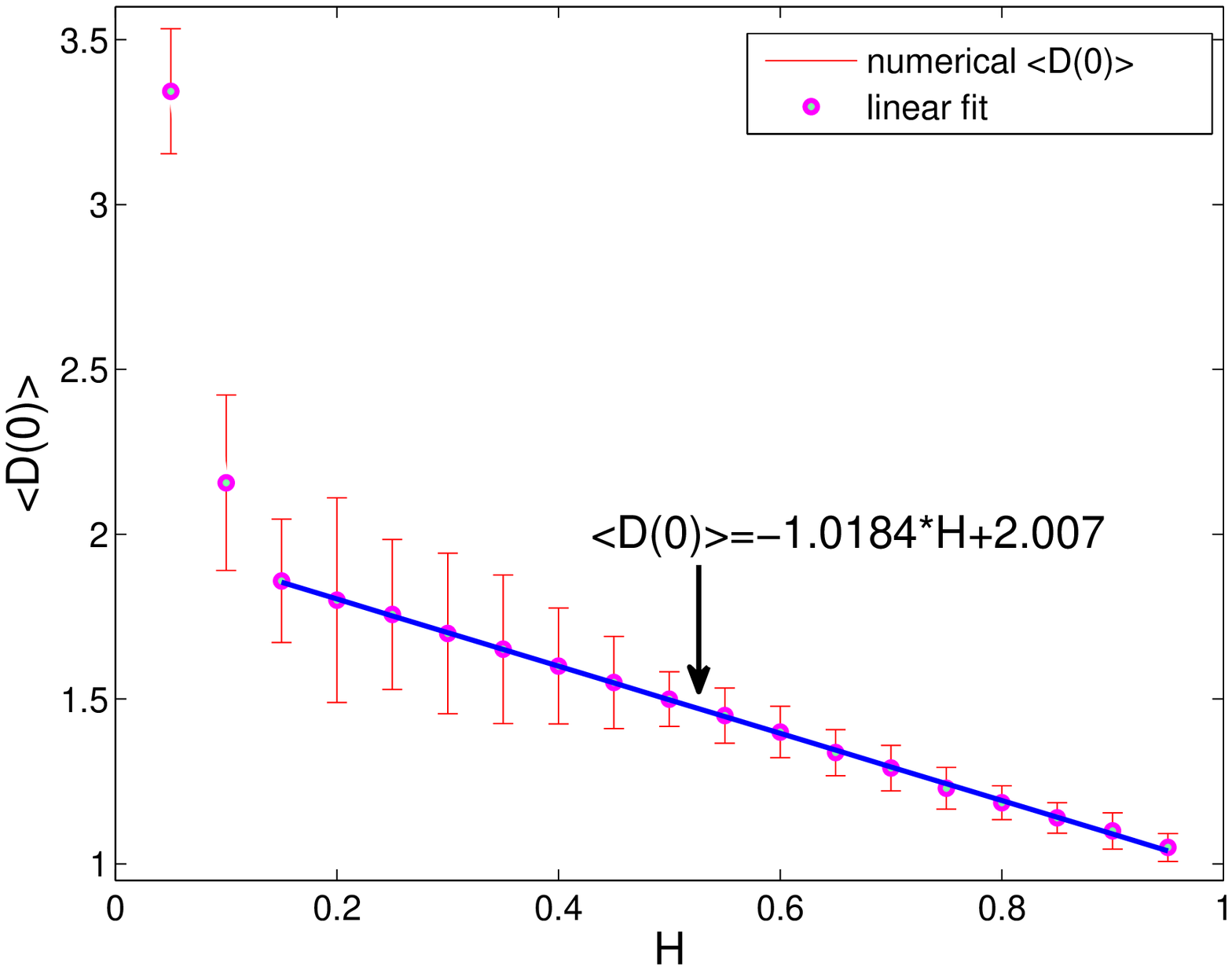}}
  \caption{The relationship between $H$ of fBm and average
  multifractality $\langle \Delta D(q) \rangle$ {\bf (Left)}, and average
  fractal dimension $\langle D(0) \rangle$ {\bf (Right)} of the associated HVGs. Here the average
  is calculated from 100 realizations, and error bars are calculated by the standard errors.}
 \end{figure}

The estimated average values of $D(0)$, $D(1)$ and $D(2)$ of HVGs constructed from fBm time series with different Hurst indices $H$ are given in Table 1.

\begin{table}
\caption{The average value of $D(0)$, $D(1)$, $D(2)$ and $\Delta D(q)$ of HVGs constructed from fBm time series with different Hurst index $H$. Here the average
is calculated from 100 realizations.}
\begin{center}
\begin{tabular}{c|c|c|c|c}
\hline
$H$ & $\langle D(0) \rangle$ &  $\langle D(1) \rangle$ &  $\langle D(2) \rangle$ & $\langle \Delta D(q) \rangle$ \\
\hline
0.20&	1.7997959&	1.5762884&	1.3987674&	1.9911429\\
0.25&	1.7566540&	1.5838531&	1.4341853&	1.6833939 \\
0.30 &  1.6987564 &  1.5615922 &  1.4340036 &  1.4104275  \\
0.35 &  1.6510399 &  1.5385958 &  1.4319206 &  1.0708452  \\
0.40 &  1.6000885 &  1.5199201 &  1.4358003 &  0.9740916  \\
0.45 &  1.5502299 &  1.4933475 &  1.4299984 &  0.7340563  \\
0.50 &  1.4997731 &  1.4635467 &  1.4240391 &  0.4474166  \\
0.55 &  1.4496548 &  1.4282816 &  1.4010555 &  0.3302060  \\
0.60 &  1.4000067 &  1.3872357 &  1.3679190 &  0.2620100  \\
0.65 &  1.3376393 &  1.3306918 &  1.3179500 &  0.2078157  \\
0.70 &  1.2907005 &  1.2839972 &  1.2721686 &  0.1899058  \\
0.75 &  1.2291436 &  1.2247421 &  1.2173406 &  0.1339882  \\
0.80 &  1.1857852 &  1.1793346 &  1.1696728 &  0.1273265  \\
0.85 &  1.1394171 &  1.1295888 &  1.1174955 &  0.1238226  \\
0.90 &  1.1000428 &  1.0834080 &  1.0638481 &  0.1914100  \\
0.95 &  1.0499715 &  1.0178998 &  0.9860908 &  0.2535463  \\
\hline
\end{tabular}
\end{center}
\end{table}

We show the relationship between the Hurst
index $H$ and the average fractal dimension $\langle D(0) \rangle$ in the right panel of Figure 3.
We can see that the average fractal
dimension $\langle D(0) \rangle$ decreases with increasing $H$. Furthermore, it
is pleasing that the curve shows a nice linear relationship:
$$\langle D(0)\rangle = 2.007 - 1.0184*H,$$
when $H\ge 0.15$,
which approximates the theoretical relationship between the Hurst
index $H$ and the fractal dimension $d$ of the graph of fBm $d = 2
- H$. Our numerical results show that the fractal dimension of the
constructed HVGs approximates closely that of the graph
of the original fBm. In other words, the fractality of the fBm is
inherited in their HVGs. This result was also reported by
Xie and Zhou \cite{XieZhou11}, where they calculated the fractal dimension of
HVGs by the simulated annealing algorithm. The functional relationships of the
average information dimension $\langle D(1) \rangle$ and the average correlation
dimension $\langle D(2)\rangle$ with the Hurst index $H$ are given in Figure 4. As
shown in Figure 4, we find that these relationships can be well
fitted by the following bi-linear functions:
$$\langle D(1) \rangle=\left\{
\begin{array}{ccc}
-0.4267*H+1.6845,~~~when~ 0.15\leq H\leq 0.5,\\
-0.9971*H+1.9758,~~~when~ 0.5\leq H\leq 0.95
\end{array}\right.$$
and
$$\langle D(2) \rangle=\left\{
\begin{array}{ccc}
-0.0049*H+1.4275,~~~when~ 0.15\leq H\leq 0.5,\\
-0.9761*H+1.9515,~~~when~ 0.5\leq H\leq 0.95.
\end{array}\right.$$

\begin{figure}[ptb]
\centerline{\epsfxsize=6.5cm \epsfbox{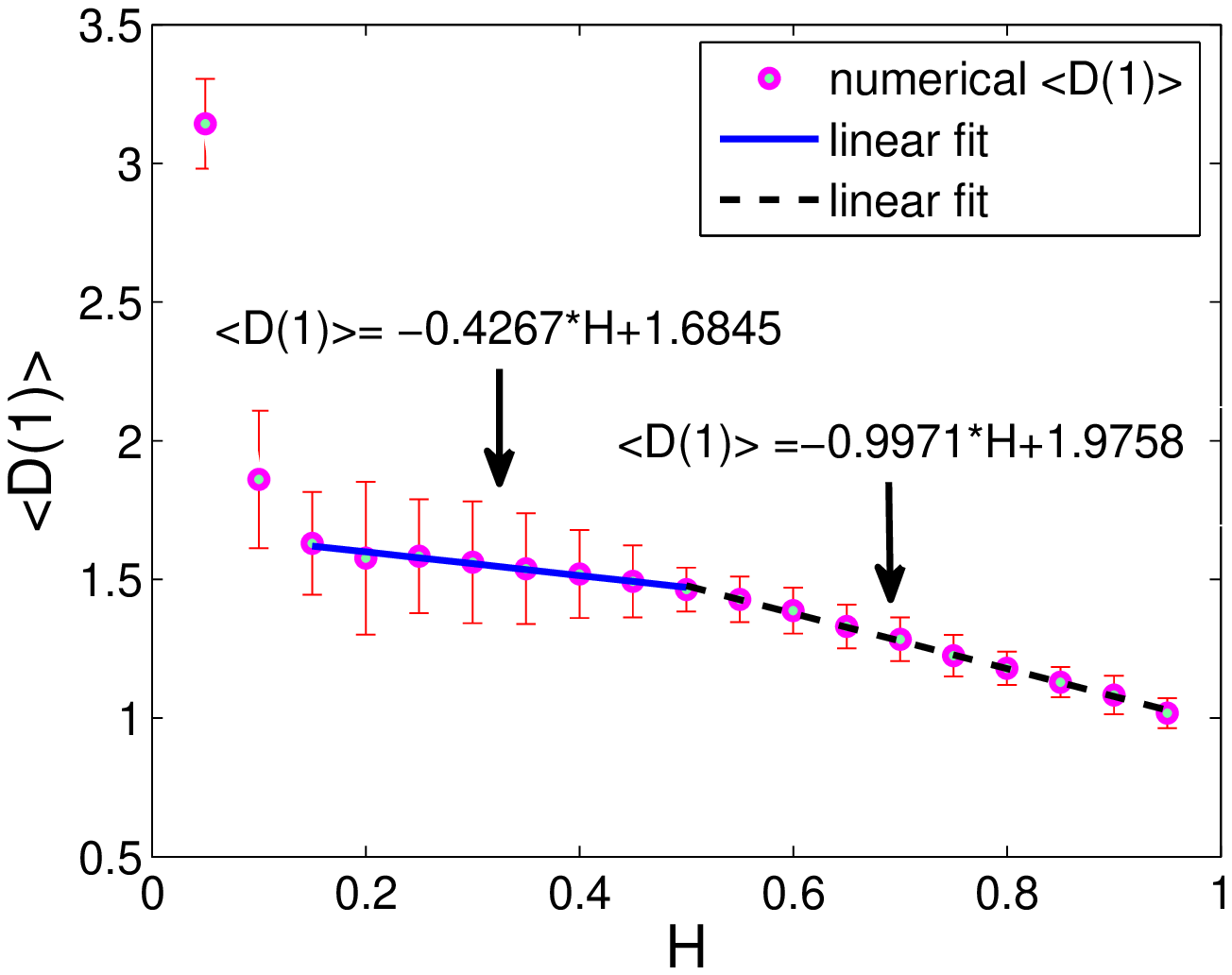}
\epsfxsize=6.5cm \epsfbox{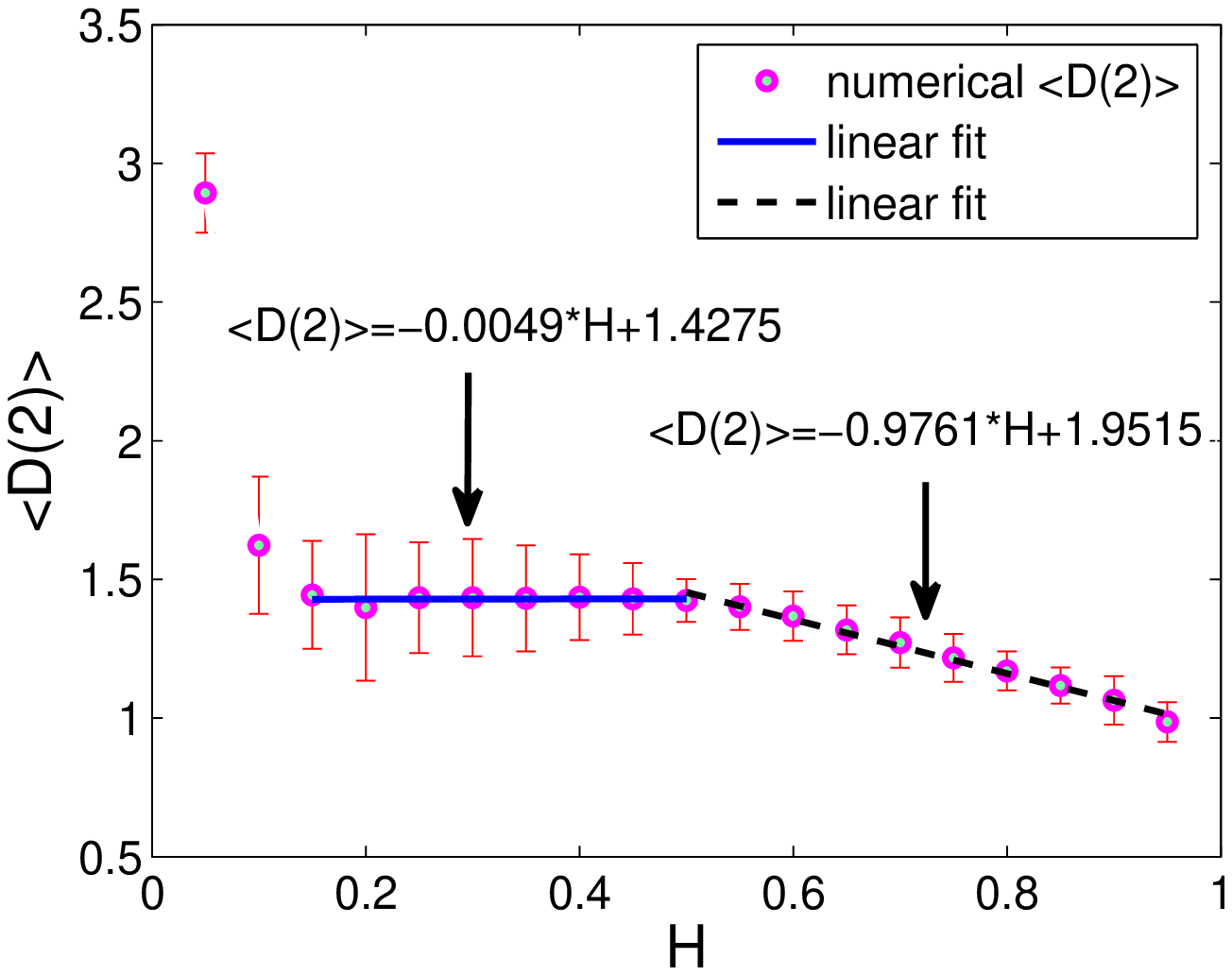}}
  \caption{The relationship between $H$ of fBm and average
  information dimension $\langle D(1) \rangle$ ({\bf Left}), and average
  correlation dimension $\langle D(2) \langle$ ({\bf Right}) of the associated HVGs. Here the average
  is calculated from 100 realizations, and error bars are calculated by the standard errors.}
 \end{figure}

Then we calculate the spectrum \cite{Chung1997,LinYau10} and energy \cite{Gutman06} for the general Laplacian operator and normalized Laplacian operator of these HVGs. One can see that all the HVGs constructed are connected. The
smallest eigenvalue $u_1$ of the general Laplacian operator $L$ and normalized Laplacian operator ${\cal L}$ of a connected graph is equal to 0.
Because the second smallest eigenvalue $u_2$ and the maximum eigenvalue $u_n$ have particular meaning,
we pay more attention to the second-smallest eigenvalue $u_2$, the third-smallest
eigenvalue $u_3$, the average maximum eigenvalue $u_n$ of these two Laplacian operators of a graph in this work.
We find that for the general Laplacian operator, the average logarithm of second-smallest eigenvalue $\langle \ln (u_2) \rangle$, the average logarithm of third-smallest eigenvalue $\langle \ln (u_3) \rangle$, and the average logarithm of maximum eigenvalue $\langle \ln (u_n) \rangle$ of these HVGs are approximately linear functions of $H$; while the average Laplacian energy $\langle E_{nL} \rangle$ is approximately a quadratic polynomial function of $H$. We show these relationships in Figure 5. For the normalized Laplacian operator, $\langle \ln (u_2) \rangle$ and $\langle \ln (u_3) \rangle$ of these HVGs approximately satisfy linear functions of $H$; while $\langle \ln (u_n) \rangle$ and $\langle E_{nL} \rangle$ are approximately a 4th and cubic polynomial function of $H$ respectively. These relationships are shown in Figure 6.

\begin{figure}[ptb]
\centerline{\epsfxsize=5.5cm \epsfbox{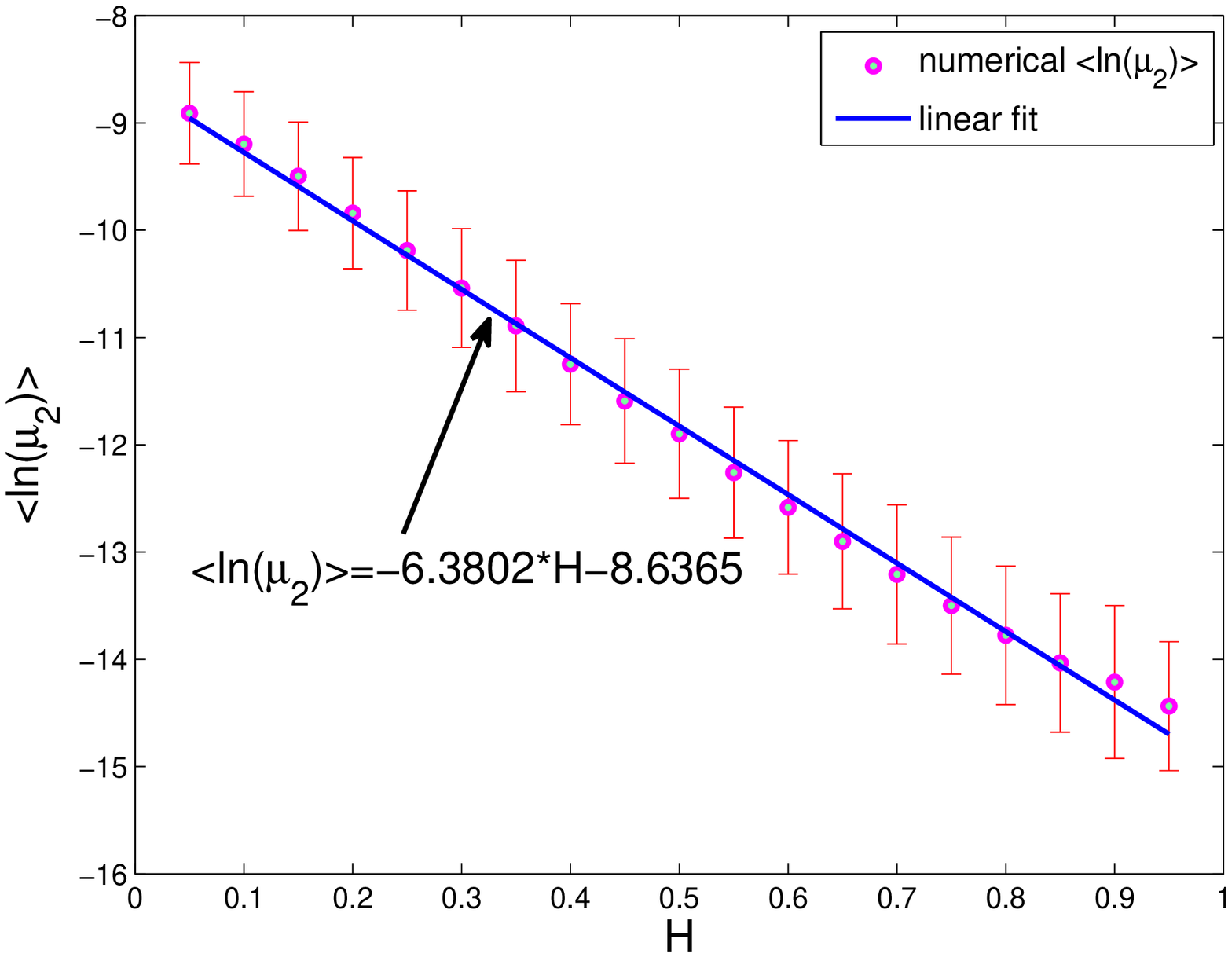}
\epsfxsize=5.5cm \epsfbox{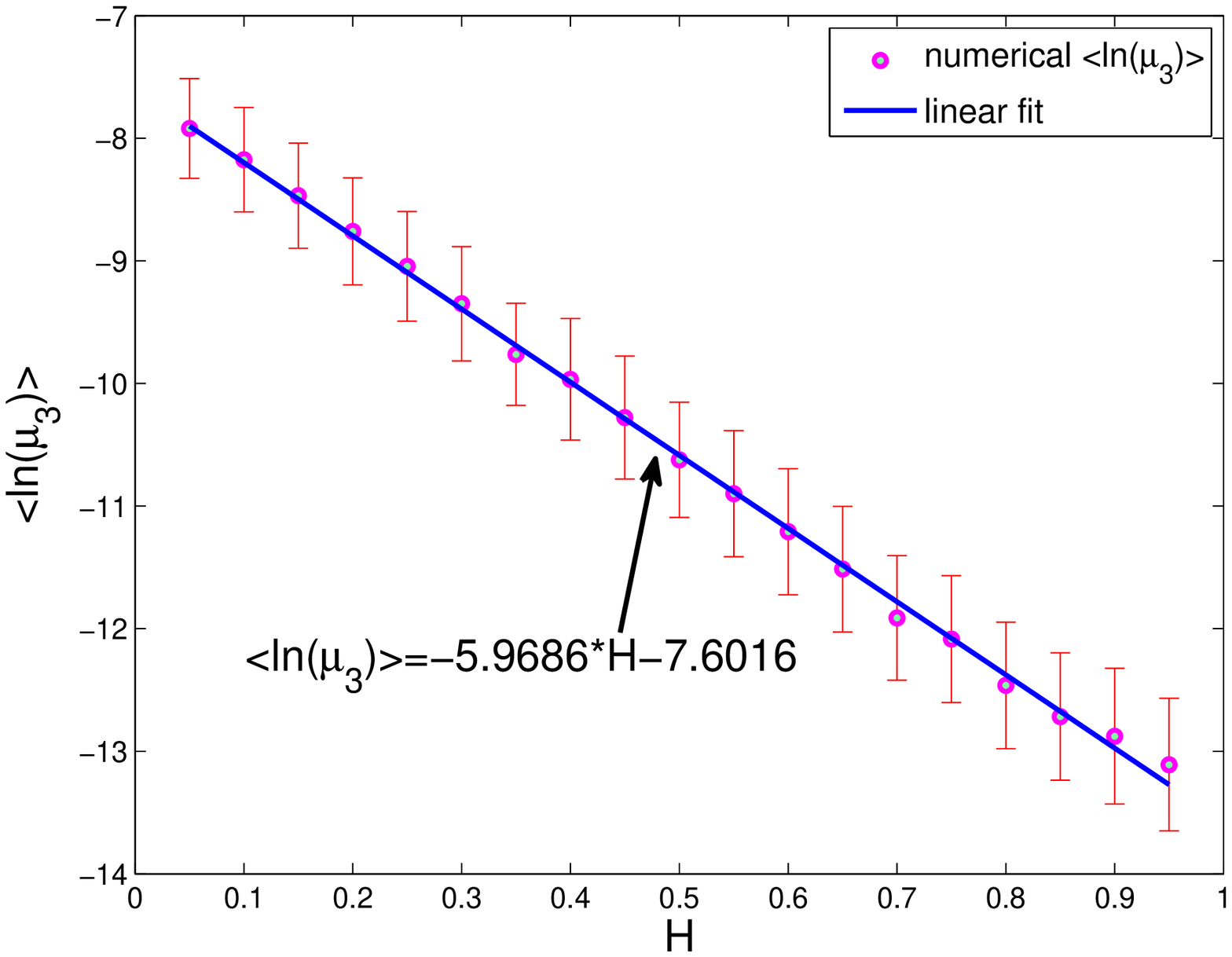}}
\centerline{\epsfxsize=5.5cm \epsfbox{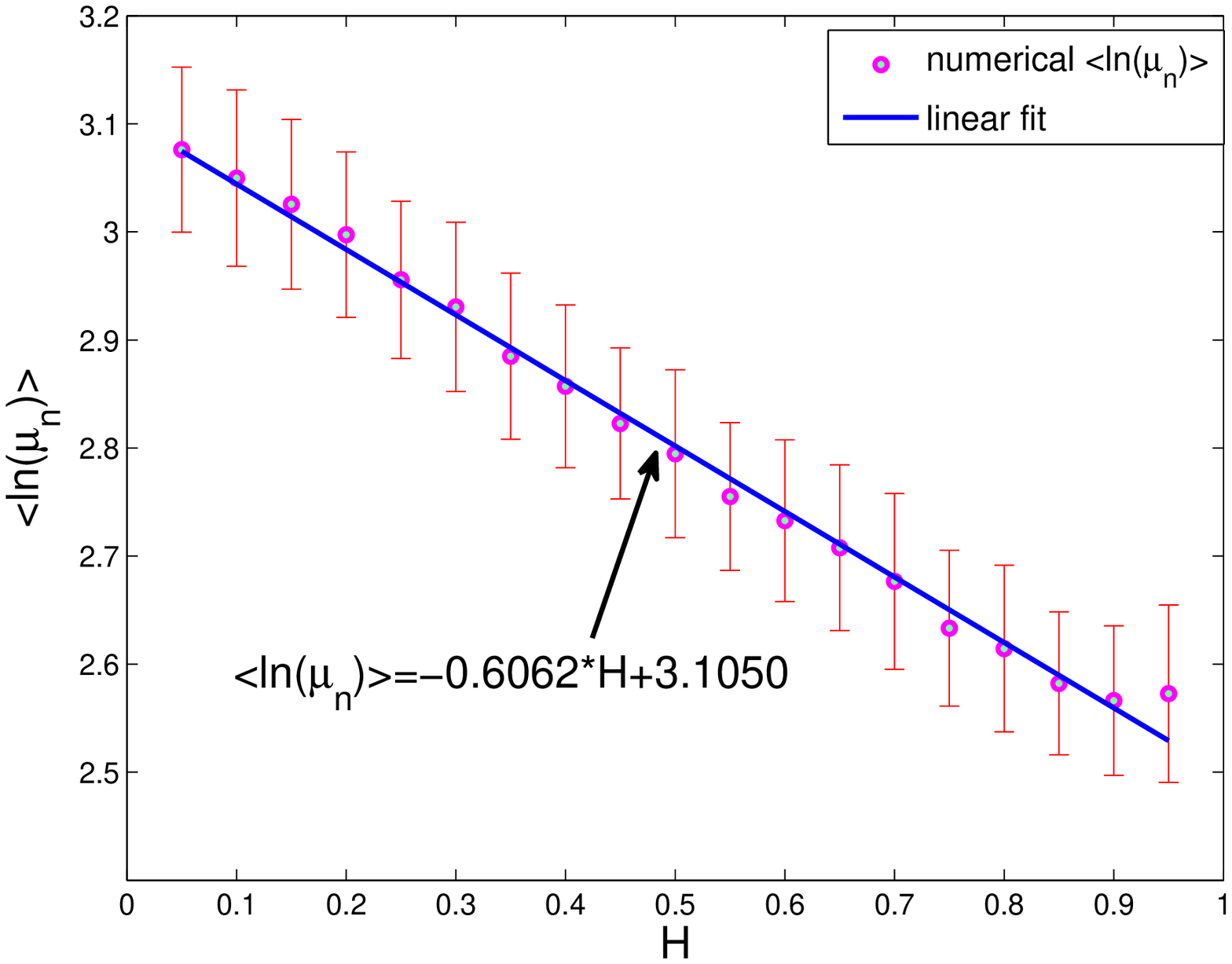}
\epsfxsize=5.5cm \epsfbox{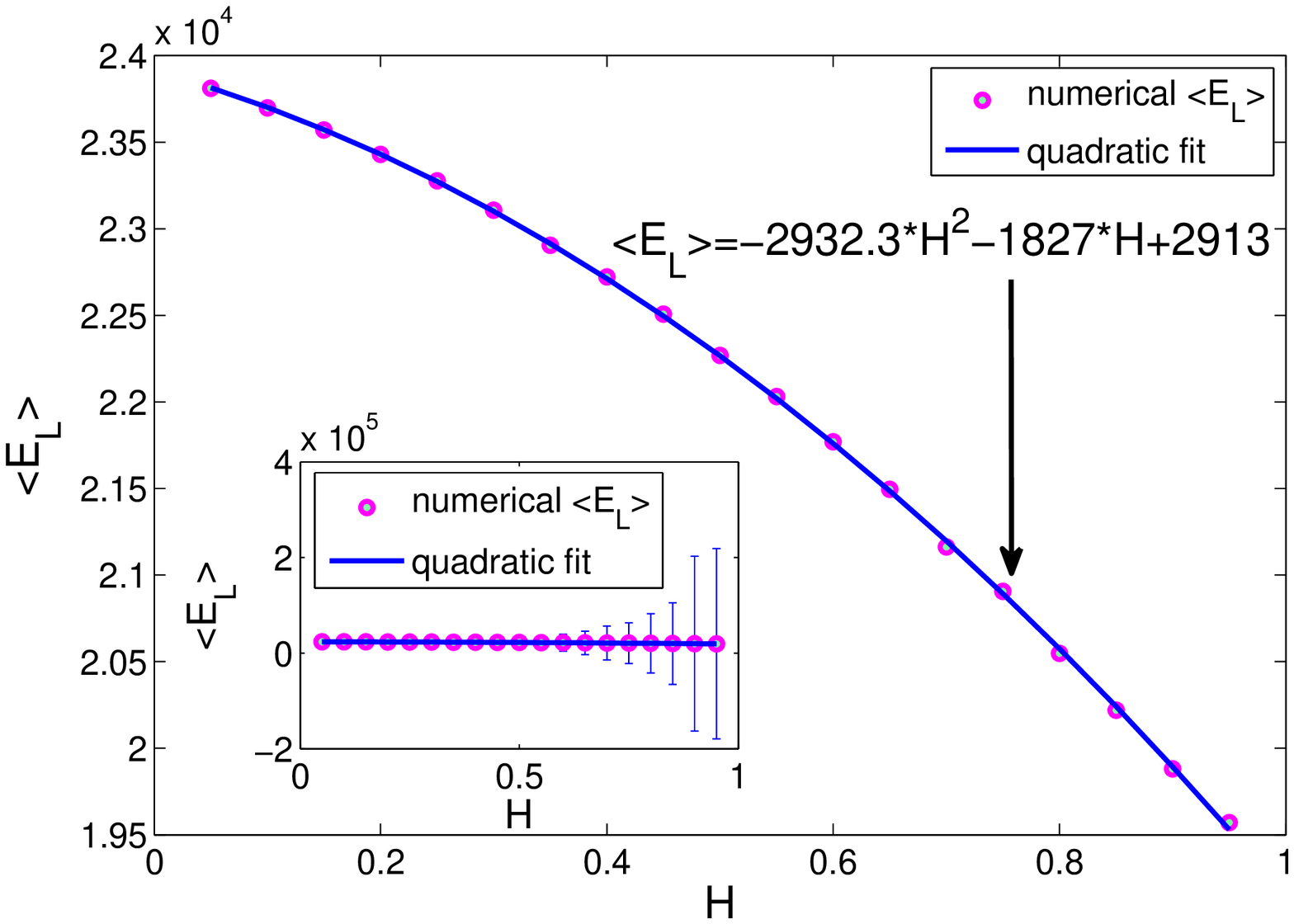}}
  \caption{The relationship between $H$ of fBm and average logarithm of
  second-smallest eigenvalue $\langle \ln(u_2) \rangle$, average logarithm of third-smallest
eigenvalue $\langle \ln(u_3) \rangle$, average logarithm of maximum eigenvalue $\langle \ln(u_n) \rangle$, and
average Laplacian energy $\langle E_{nL} \rangle$ of these HVGs for general Laplacian operator. Here the average
  is calculated from 100 realizations, and error bars are calculated by the standard errors.}
 \end{figure}

\begin{figure}[ptb]
\centerline{\epsfxsize=5.5cm \epsfbox{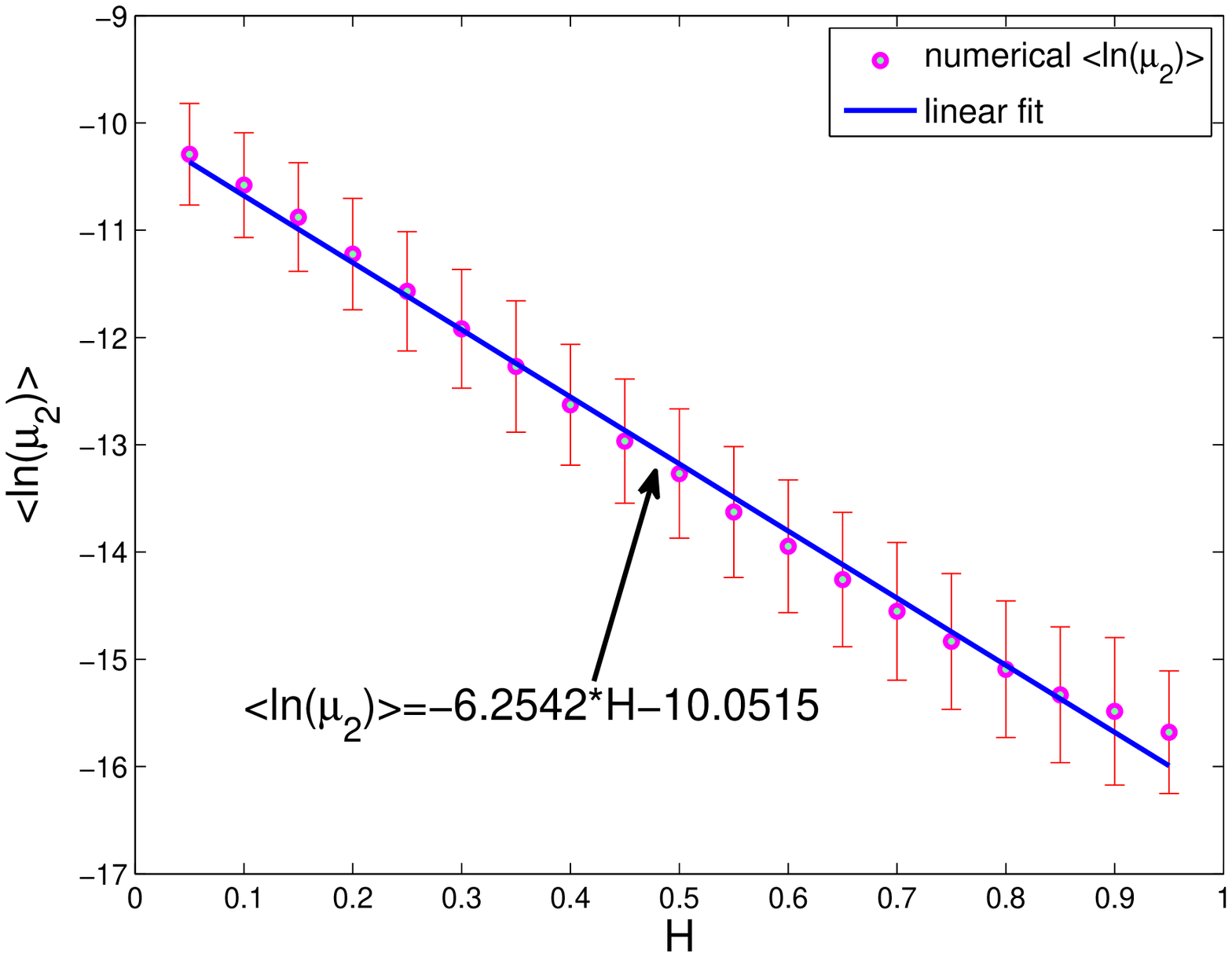}
\epsfxsize=5.5cm \epsfbox{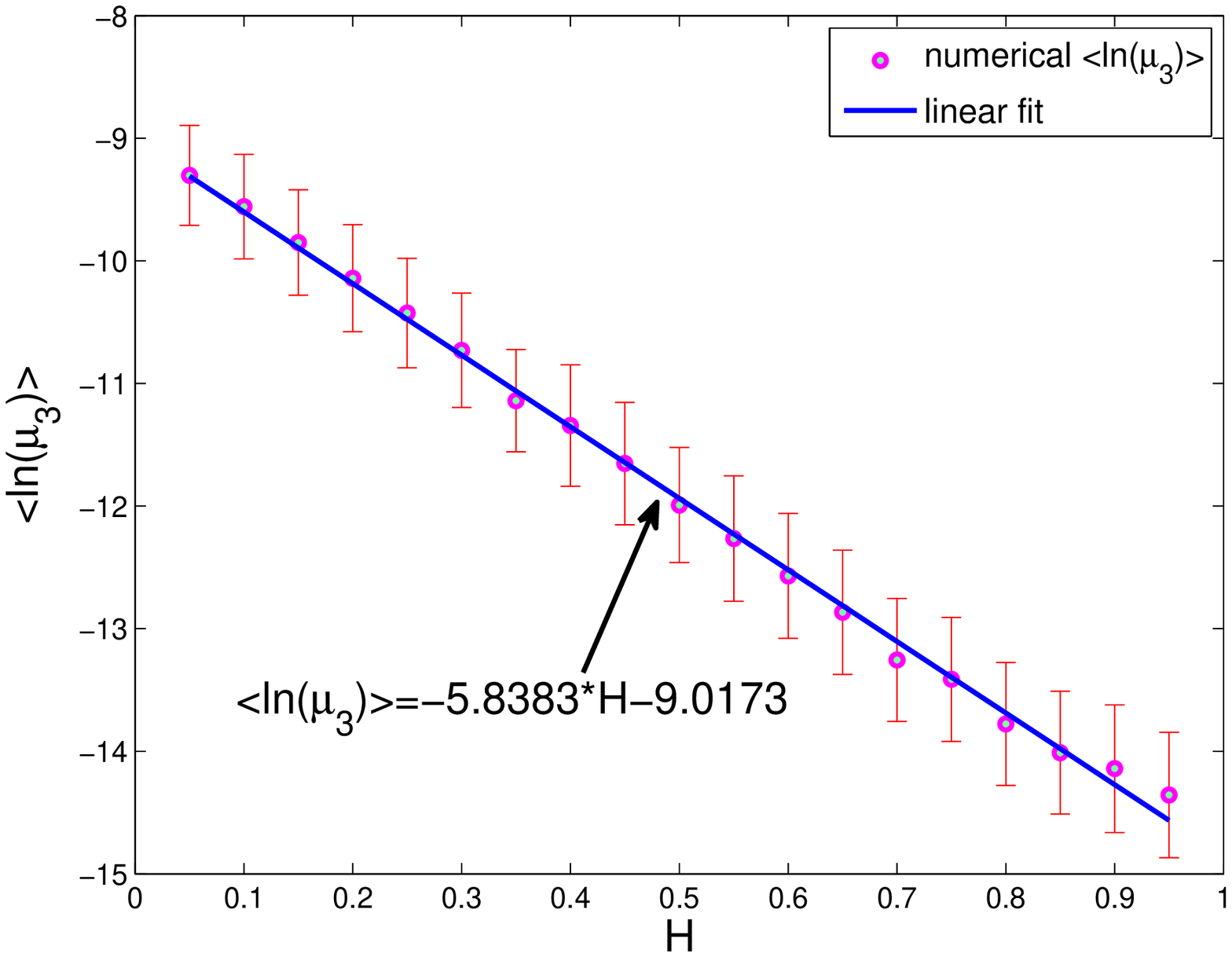}}
\centerline{\epsfxsize=5.5cm \epsfbox{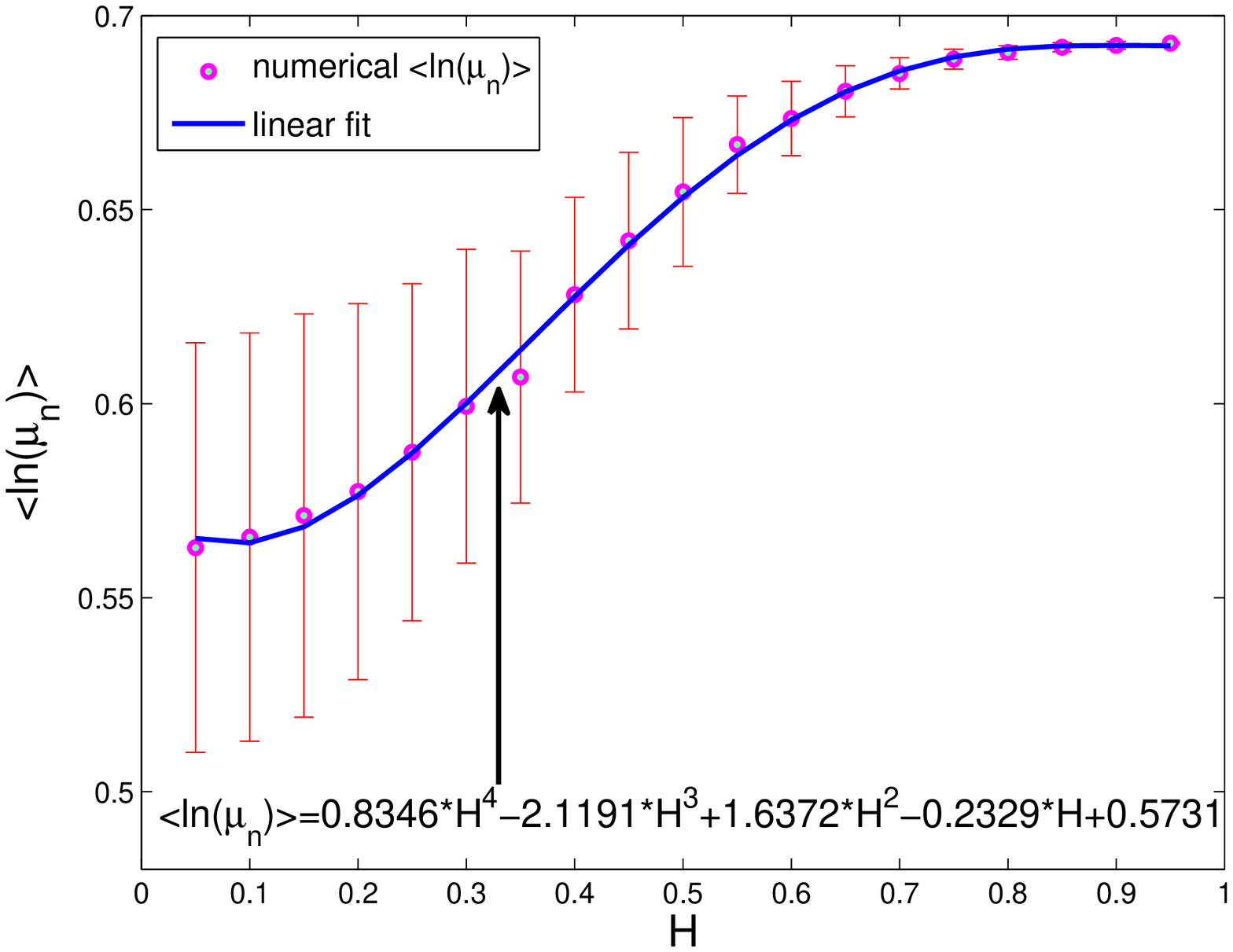}
\epsfxsize=5.5cm \epsfbox{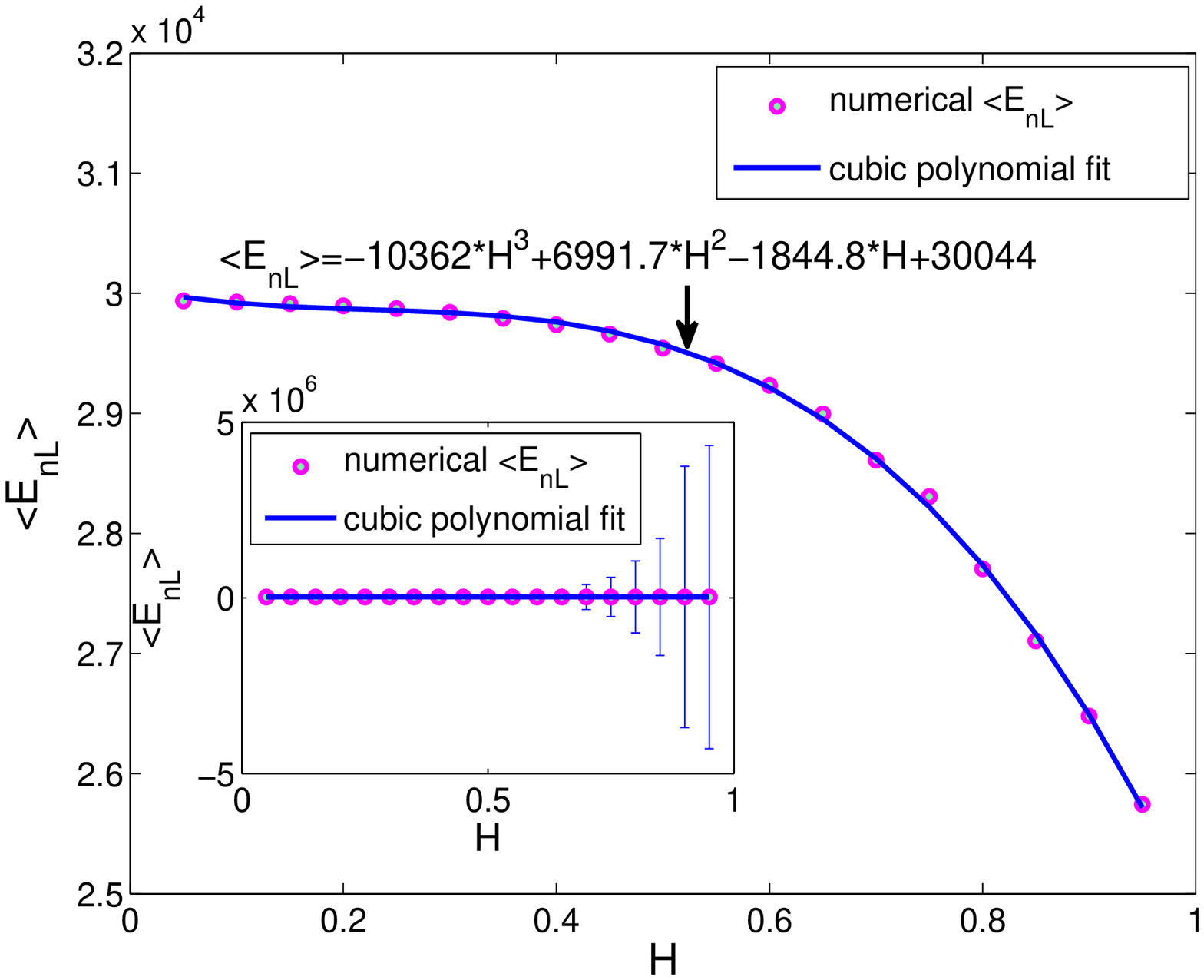}}
  \caption{The relationship between $H$ of fBm and average logarithm of
  second-smallest eigenvalue $\langle \ln(u_2) \rangle$, average logarithm of third-smallest
eigenvalue $\langle \ln(u_3) \rangle$, average logarithm of maximum eigenvalue $\langle \ln(u_n) \rangle$, and
average Laplacian energy $\langle E_{nL} \rangle$ of these HVGs for normalized Laplacian operator. Here the average
  is calculated from 100 realizations, and error bars are calculated by the standard errors.}
 \end{figure}

From the above results, we conclude that the
inherent nature of the time series affects the structure
characteristics of the associated networks and the dependence
relationships between them appear retained. Our work supports that
complex networks provide a suitable and effective tool to perform time
series analysis.

\section*{Acknowledgments}
This project was supported by the Natural Science Foundation of
China (Grant no. 11371016), the Chinese Program for
Changjiang Scholars and Innovative Research Team in University
(PCSIRT) (Grant No. IRT1179).

\end{document}